\documentstyle[aps,epsf]{revtex}  

\newskip\zatskip \zatskip=0pt plus0pt minus0pt
\def\matth{\mathsurround=0pt}
\def\gtrsim{\mathrel{\mathpalette\atversim>}}
\def\lstsim{\mathrel{\mathpalette\atversim<}}
\def\atversim#1#2{\lower0.7ex\vbox{\baselineskip\zatskip\lineskip
  \zatskip
  \lineskiplimit 0pt\ialign{$\matth#1\hfil##\hfil$\crcr#2\crcr\sim
  \crcr}}}

\begin{document}        

\baselineskip 14pt
\begin{titlepage}

\hspace*{\fill}\parbox[t]{4cm}{
MSUHEP-90416\\
April 16, 1999\\
hep-ph/9904368}

\vspace{1.cm}
\begin{center}
{\Large\bf Status of the BFKL Resummation Program\footnote{
Talk given at DPF '99, Los Angeles, CA, 1999.}} \\
\vspace{1.cm}

{Carl R. Schmidt}\\
\vspace{.2cm}
{\sl Department of Physics and Astronomy\\
Michigan State University\\
East Lansing, MI 48824, USA}\\

\vspace{1.cm}

\begin{abstract}
I discuss the calculation of the next-to-leading logarithmic (NLL) 
corrections to the BFKL resummation, as well as some of the issues 
that arise in this formalism at NLL.  In particular I consider the 
large size and apparent instability of the corrections, and I 
address some of the attempts to understand and tame them.
\hspace*{\fill}
\end{abstract}

\end{center}
 \vfil

\end{titlepage}

\title{Status of the BFKL Resummation Program}
\author{Carl R. Schmidt}
\address{Department of Physics and Astronomy, 
Michigan State University, East Lansing, Michigan 48824}
%
\maketitle              

\begin{abstract}        

I discuss the calculation of the next-to-leading logarithmic (NLL) 
corrections to the BFKL resummation, as well as some of the issues 
that arise in this formalism at NLL.  In particular I consider the 
large size and apparent instability of the corrections, and I 
address some of the attempts to understand and tame them.

\
\end{abstract}   	

\section{Introduction to NLL BFKL}      

Early last year, after many years of hard work involving many 
participants, Fadin and Lipatov \cite{nll} presented the NLL 
corrections to the BFKL equation.  Since then, there has been much 
lively discussion of the interpretation of these corrections.  In 
this talk I will discuss some of the results of this activity.
I will not address phenomenology here, but some applications of the 
BFKL resummation are 
inclusive dijet production at large rapidity separation $ 
y=\ln\hat s/(p_{1\perp}p_{2\perp})$, forward jet production in 
deep-inelastic scattering (DIS), DIS structure functions at small-$x$, 
and others.

In order to discuss the NLL corrections it is useful 
to consider first the solution to the BFKL equation at LL \cite{FKL}. 
The BFKL equation is used to resum all powers of 
$\alpha_{s}\log(\hat s)$ in the cross section.  This leads to the 
familiar prediction that at very high energies the cross section
scales as a power of the energy:
\begin{equation}
	\hat\sigma\ \approx e^{Ay} \approx\ 
	{\hat s}^{A}\ .
	\label{saddle}
\end{equation}
The quantity $(1+A)$ is often referred to as the BFKL Pomeron 
intercept.
This scaling behavior is obtained from the solution to 
the BFKL equation, which is given at LL by the integral
\begin{equation}
f(y,p_{1\perp},p_{2\perp})\ =\ {1\over 2\pi i}
\int_{1/2-i\infty}^{1/2+i\infty} d\gamma\, 
(p_{1\perp}^2)^{\gamma-1}\,(p_{2\perp}^{2})^{-\gamma}
e^{\bar\alpha_{s}\chi^{(0)}(\gamma)y}\ ,\label{sold}
\end{equation}
where $\bar \alpha_{s}=\alpha_{s}N_{c}/\pi$, and we have performed an 
azimuthal average over the transverse momenta for convenience.  The 
function 
\begin{equation}
	\chi^{(0)}(\gamma)\ =\ 2\psi(1)-\psi(\gamma)-\psi(1-\gamma)
	\label{eig}
\end{equation}
is the eigenvalue of the LL BFKL kernel, where $\psi$ is the  
logarithmic derivative of the gamma function.  Performing the 
integral in the saddle-point approximation leads to the exponential 
rise in the cross section (\ref{saddle}) 
with $A=\bar\alpha_{s}\chi^{(0)}({1\over2})=4\bar\alpha_{s}\ln2$. 

At the heart of the BFKL equation, which is used to derive (\ref{sold}),
 is the kernel
$K(p_{1\perp},p_{2\perp})$, which is an 
integral operator in transverse momentum space that is used to build 
up the BFKL ladder.  The contributions to the kernel are shown, 
schematically, at LL and NLL in Fig.~1.  
At LL each application of the kernel adds one more factor of 
${\cal O}(\alpha_{s}\Delta y)$ to the resummation.   It is composed of 
two types of contributions.  The first type corresponds to an emission 
of a real gluon, in the approximation that it is widely separated in 
rapidity from any other emissions (known as multi-Regge kinematics).  
The factor of $\Delta y$ just comes 
from the integration over the rapidity of this gluon.  The second type 
corresponds to the virtual contributions which are enhanced by the 
logarithmic factor $\Delta y$.  The simplest contribution of this 
second type can be found by considering the one-loop 
corrections to $gg\rightarrow gg$ scattering.

At NLL one also includes terms of ${\cal O}(\alpha_{s}^{2}\Delta y)$
with each application of the kernel.
They consist of three types, corresponding to:  the emission of 
two gluons nearby in rapidity, the virtual correction to the emission
of one gluon in the multi-Regge kinematics, and the subleading 
purely-virtual corrections.  This last contribution can be found by 
considering 
the two-loop corrections to $gg\rightarrow gg$ scattering.
It is the calculation of these three contributions which took many 
years and many papers to sort out the technical details\footnote{A 
list of references can be found in ref.~\cite{schmidt}, but with no 
guarantee of completeness.}  Although the full kernel has not been 
checked in a completely independent manner, many of the pieces of the 
calculation have received independent confirmation.  Two particularly 
significant checks are the 
calculation of the virtual correction to the gluon emission in 
multi-Regge kinematics \cite{ddsii}, and the compilation of the three 
NLL terms into a single kernel with the cancellation of all collinear and 
soft singularities \cite{cc}.

\begin{figure}[t]	
\centerline{\epsfxsize 4.0 truein \epsfbox{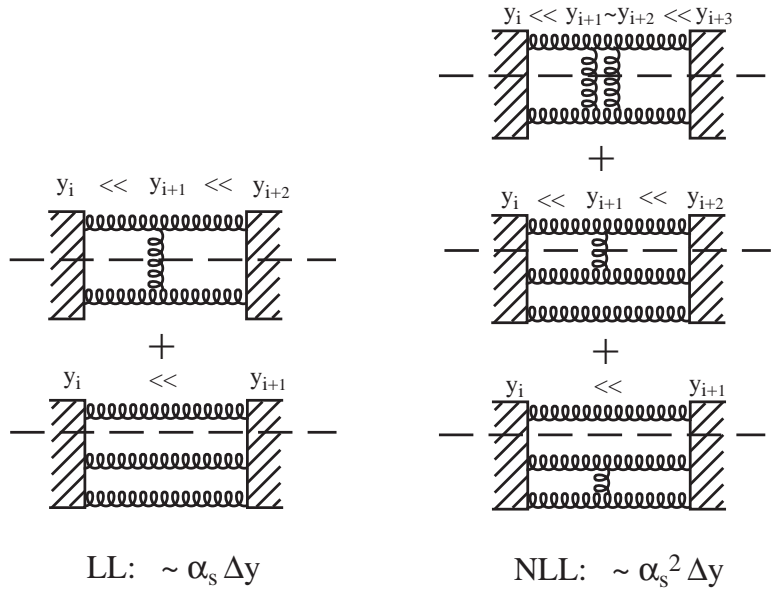}}
\vskip 0.0 cm
\caption[]{
\label{kernel}
\small Contributions to the BFKL kernel at LL and NLL.  Gluons that
cross the dashed line correspond to real emission.}
\end{figure}

The final result of this calculation is usually presented by applying 
the kernel to the LL eigenfunctions, with azimuthal averaging, yielding
\begin{eqnarray}
	\int d^{2}p_{2\perp}K(p_{1\perp},p_{2\perp})
	(p_{2\perp}^{2}/p_{1\perp}^{2})^{\gamma-1}\ &=& \ \
	\bar\alpha_{s}(\mu)\,\chi(\gamma)\nonumber\\
	&=&\ \ \ 
	\bar\alpha_{s}(\mu)\,
	\chi^{(0)}(\gamma)\left[
	1-\bar\alpha_{s}(\mu)b_{0}
	\ln(p_{1\perp}^{2}/\mu^{2})\right]\nonumber\\
	&&
	+\ \bar\alpha_{s}^{2}(\mu)\chi^{(1)}(\gamma)\ ,
	\label{nll}
\end{eqnarray}
where $b_{0}= {11/12}-{n_{f}/(6N_{c})}$ and $\mu$ is the
 $\overline{\rm MS}$ renormalization scale.
The NLL correction has been separated into two terms.  The first term 
depends on the scale $p_{1\perp}$ and is associated with the running 
of the coupling in the LL kernel:  
$\alpha_{s}(\mu)\rightarrow \alpha_{s}(p_{1\perp})$.  
The second term, $\bar\alpha_{s}^{2}\chi^{(1)}(\gamma)$, is independent 
of scale and contains the remainder of the NLL corrections \cite{nll}.

\section{Problems at NLL.}

After completion of the NLL corrections to the BFKL kernel, several 
issues with the NLL solution quickly became apparent.  Depending on 
one's point of view, these may even signify critical problems 
for the entire BFKL resummation program.  
Roughly speaking, they can be separated into 
issues associated with the running coupling term and issues associated 
with the scale-invariant term.  Although my main focus will be 
on the scale-invariant term, I briefly 
touch on the topic of the running coupling term in NLL BFKL.

\subsection{Running coupling problems.}

The issues that arise with the running of the coupling in BFKL were 
not entirely unanticipated.  They are related to the fact 
that the emission of each real gluon causes the momentum carried 
down the BFKL ladder to 
diffuse as one moves away from the starting rapidity.  It 
can diffuse to larger values or to smaller values; however, if it 
diffuses below values around $\Lambda_{QCD}$ then nonperturbative 
effects become important, and one can no longer make unambiguous 
predictions from the perturbative BFKL resummation.  This issue was 
known even before the NLL corrections were completed, although it can
be ignored in LL BFKL.  This is because logarithms 
involving two transverse scales do not arise until NLL, and so 
LL BFKL is calculated at fixed coupling.  At NLL, however,
this issue can no longer be swept under the rug.

The effects of the running coupling term in the NLL BFKL solution 
have been considered in several papers.  Armesto, Bartels, and Braun 
\cite{bartels} considered the modification of the eigenvalues and the 
eigenfunctions due to the NLL terms in the kernel.  They found that, due 
to the running coupling term in the kernel, the NLL eigenvalues can take 
on any value along the real axis.  This is unlike at LL, where the 
eigenvalues (\ref{eig}) have a maximum and the equation 
(\ref{sold}) is well-defined.  Thus, the interpretation of the NLL 
corrections is not entirely well-defined in this approach.
Correspondingly, the NLL eigenfunctions 
contain pieces displaying non-perturbative behavior.

The effects of the running coupling term in the NLL BFKL solution 
were also included through a different approach by Kovchegov and 
Mueller \cite{mueller}.  They obtained a NLL solution by explicitly 
iterating the NLL corrections to the kernel, starting from the LL solution 
evaluated in the saddle point approximation.  They found that the 
running coupling term leads to a non-Regge behavior in the energy 
dependence of the cross section. (This was also shown in
 ref.~\cite{levin}).  This 
non-Regge behavior is exhibited as a term of the form $\alpha_{s}^{5}y^{3}$ 
in the exponential of eq.~(\ref{saddle}) at high energies.  In 
addition, they showed that the nonperturbative effects from resumming
the logarithms in the running coupling should be small
as long as $14\zeta(3)b_{0}^{2}\bar \alpha_{s}(\mu)^{3}y\lstsim1$.

From these results, it appears that for some region of kinematics the 
nonperturbative effects should be small; however, the proper way to 
deal with them at NLL is not yet completely clear.

\subsection{Scale-invariant problems.}

Whereas the problems at NLL due to the running coupling were 
anticipated to some degree, the problems due to the scale-invariant 
term were a big surprise.  The first indication of this problem was 
seen immediately by Fadin and Lipatov.  The corrections to the 
leading eigenvalue  are large and negative!  If we ignore the running 
coupling term, we obtain 
\begin{equation}
	\bar\alpha_{s}\chi(\mbox{$1\over2$})\ =\ 
	2.77\bar\alpha_{s}-18.34\bar\alpha_{s}^{2}\ ,
	\label{nllchi}
\end{equation}
for three active flavors.  At the not-unreasonable value 
of $\alpha_{s}=0.16$ the NLL corrections exactly cancel the LL term, 
while for larger values of $\alpha_{s}$ the eigenvalue becomes negative.
Naively, this would indicate that the BFKL Pomeron intercept also 
becomes negative, leading to a cross section that decreases, rather 
than increases, as a power of the energy.

Of course, this interpretation relies on the saddle-point evaluation 
of the NLL generalization of the BFKL solution (\ref{sold}).  Upon 
closer analysis Ross \cite{ross} showed that the NLL eigenvalue 
function $\chi(\gamma)$ no longer has a maximum at $\gamma={1\over2}$, but 
has a minimum with two maxima occuring symmetrically on either side 
of this point\footnote{The standard procedure in these analyses is to 
modify the LL eigenfunctions used in eq.~(\ref{nll}) in order to make the
eigenvalues manifestly symmetric under $\gamma\rightarrow1-\gamma$, 
following
ref.~\cite{nll}.}.  Performing a higher-order expansion of $\chi(\gamma)$, 
Ross found a smaller correction to the BFKL Pomeron intercept.
However, the solution he obtained was not positive definite.  It 
contained oscillations as one varied $p_{1\perp}$ and $p_{2\perp}$.
This led Levin \cite{levin} to declare that NLL BFKL has a serious 
pathology.

One might wonder whether the approximate evaluation of the integral 
performed by Ross is adequate at this stage.  Perhaps an exact evaluation 
is necessary.  However, negative cross sections have also arisen when 
the resummed small-$x$ anomalous dimensions, obtained from the NLL BFKL 
solution, were used to study DIS scattering at 
small-$x$ \cite{ball}.  In any event the NLL corrections to the 
BFKL solution are large, leading one to question the stability and 
applicability of the BFKL resummation procedure in general.

\section{Attempts to Fix/Understand the large NLL Corrections.}

In this section I will discuss several attempts to understand the origin 
of the large NLL corrections and to control them.  The first two 
proposals, although very different in implementation, both can be 
traced to correlations that arise when two neighboring gluons are 
emitted close to each other in rapidity \cite{bo}.  Essentially, the
LL BFKL equation greatly overestimates the contribution of this 
collinear 
configuration because of the lack of ordering in transverse momentum.

The first proposal by Salam \cite{salam} was to resum the double transverse 
logarithms of the form 
$\bar\alpha_{s}\ln^{2}(p_{1\perp}^{2}/p_{2\perp}^{2})$.  This idea 
is based on the studies of Camici and 
Ciafaloni \cite{cc} on the energy-scale dependence of NLL BFKL.   
Instead of choosing the symmetric rapidity 
$y=y_{2}-y_{1}=\ln\hat s/(p_{1\perp}p_{2\perp})$ as the large  
logarithm to resum, one could equally well have chosen $y^{+}=
\ln x^{+}_{2}/x^{+}_{1}=\ln \hat s/p_{1}^{2}$ 
or $y^{-}=\ln x^{-}_{1}/x^{-}_{2}=
\ln \hat s/p_{2}^{2}$, where $x^{\pm}_{i}$ is the momentum fraction along 
the positive or negative light-cone for the emitted gluon $i$.
 Although these choices are all 
equivalent at LL, at NLL a change in the logarithm produces a change 
in the NLL kernel and can introduce the double transverse logarithms.

Motivated by DGLAP-type resummation \cite{dglap} one finds that the 
appropriate choice is to resum $y^{+}$ when $p_{1\perp}^{2}\gg 
p_{2\perp}^{2}$ and $y^{-}$ when $p_{2\perp}^{2}\gg 
p_{1\perp}^{2}$.  In refs.~\cite{cc} and \cite{nll} it was shown 
that changing 
from $y^{+}$ to $y$ shifts the NLL eigenvalue by terms with 
$1/\gamma^{3}$ singularities.  Similarly, changing 
from $y^{-}$ to $y$ shifts the NLL eigenvalue by terms with 
$1/(1-\gamma)^{3}$ singularities.  Both the $1/\gamma^{3}$ and the 
$1/(1-\gamma)^{3}$ singularities can be identified in 
$\chi^{(1)}(\gamma)$, and methods for resumming these singularities 
were given in ref.~\cite{salam}.  

Results of this resummation of double transverse logarithms are shown in 
Fig.~2, where the leading eigenvalue $\bar\alpha_{s}\chi({1\over2})$ and 
its second derivative are plotted as a function of $\bar\alpha_{s}$.  
The different schemes 1--4 give some measure of the ambiguity in this 
resummation procedure.  In general the eigenvalue is found to be 
positive after resummation, although less than at LL.  In addition the 
point $\gamma={1\over2}$ remains a maximum over a wider range of 
$\alpha_{s}$, especially in schemes 3 and 4.  

The physical implications of this proposed solution 
can be seen by further investigating the relation between 
resummation in $y^{\pm}$ and $y$.  When $p_{1\perp}^{2}\gg
p_{2\perp}^{2}$, the resummation in $y^{+}$ requires the ordering
$x^{+}_{2}>x^{+}_{1}$.
  Translating back into the symmetric variable 
$y$, this implies $y_{2}-y_{1}>\ln(p_{1\perp}/p_{2\perp})$.  Similarly,  
when $p_{2\perp}^{2}\gg 
p_{1\perp}^{2}$, the resummation in $y^{-}$  requires the ordering
$x^{-}_{1}>x^{-}_{2}$, implying
$y_{2}-y_{1}>\ln(p_{2\perp}/p_{1\perp})$.  These constraints hold for 
any two successively emitted gluons.  Therefore, the resummation of 
the double transverse logarithms corresponds to imposing a 
$p_{\perp}$-dependent cut, 
$y_{i+1}-y_{i}>|\ln(p_{i\perp}/p_{i+1\perp})|$, 
 on the separation in rapidity between the 
neighboring gluons.

This leads to the second proposal \cite{schmidt} (first suggested in 
\cite{bo} and \cite{liptalk}) for dealing with the large corrections 
to BFKL at NLL.  It is to introduce explicitly a rapidity separation 
parameter $\Delta$ into the BFKL equation, enforcing the condition 
$y_{i+1}-y_{i}>\Delta$, where $\Delta$ is assumed to be much less than 
the total rapidity interval $y$.  This parameter can be included 
systematically 
at any order in the resummation, and it plays a role for the 
rapidity resummation similar to the role played by the $\overline{\rm MS}$
renormalization scale $\mu$ for the resummation of logarithms in the 
running coupling.  A change in $\Delta$ shifts pieces of the 
calculation between LL and NLL, such that any differences are always 
next-to-next-to-leading logarithm
(NNLL).  Thus, the dependence on $\Delta$ can be regarded as an estimate 
of the uncertainty due to NNLL corrections.

\begin{figure}[t]	
\vskip-1.0in
\centerline{\epsfxsize 5.5 truein
\epsfbox{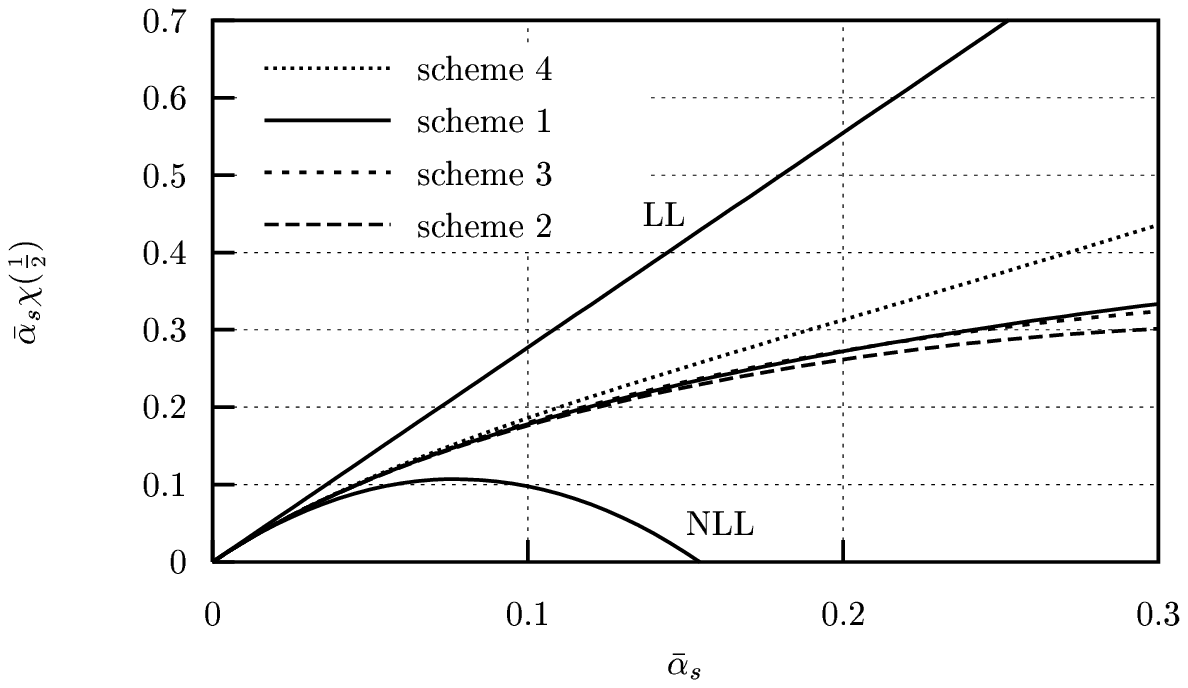}\hskip-2.2in
\epsfxsize 5.5 truein
\epsfbox{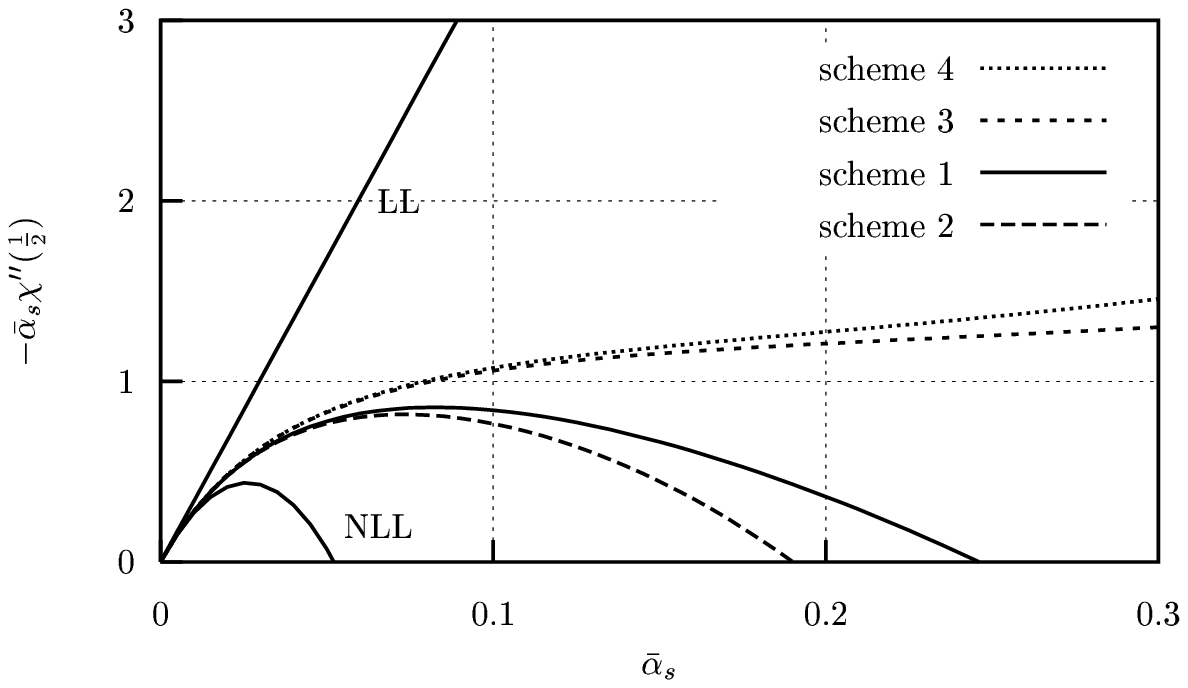}}
\vskip-4.2in
\caption[]{
\label{salamFigure}
\small The result of resumming double transverse logarithms on $\chi$ 
and its second derivative, from 
Ref.~\cite{salam}.}
\end{figure}

\vskip-0.2in
\begin{figure}[t]	
\centerline{\epsfxsize 4.0 truein \epsfbox{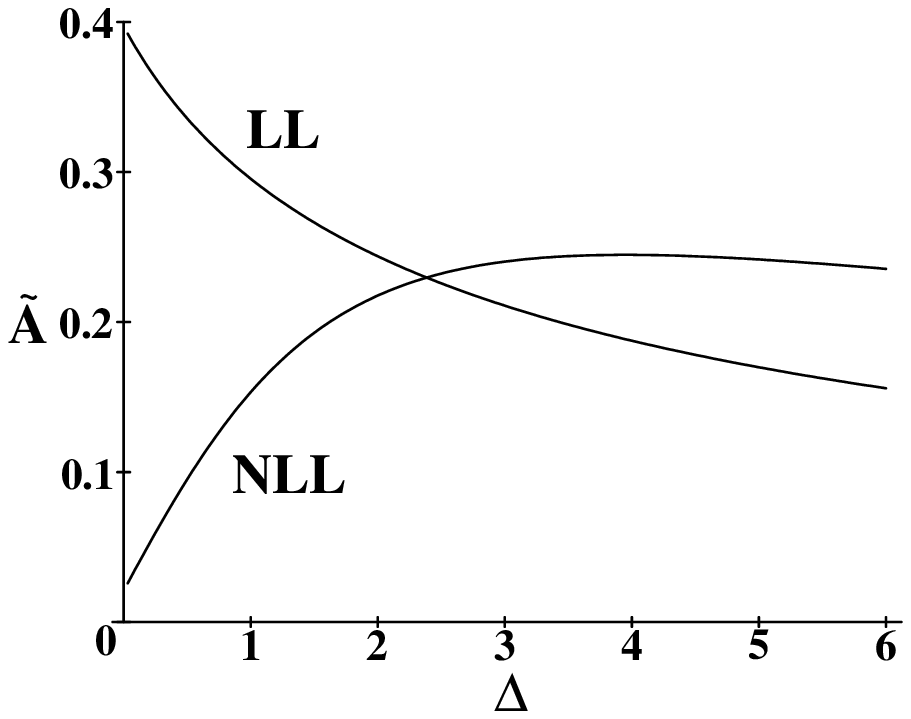}\hskip-0.8in
\epsfxsize 4.0 truein \epsfbox{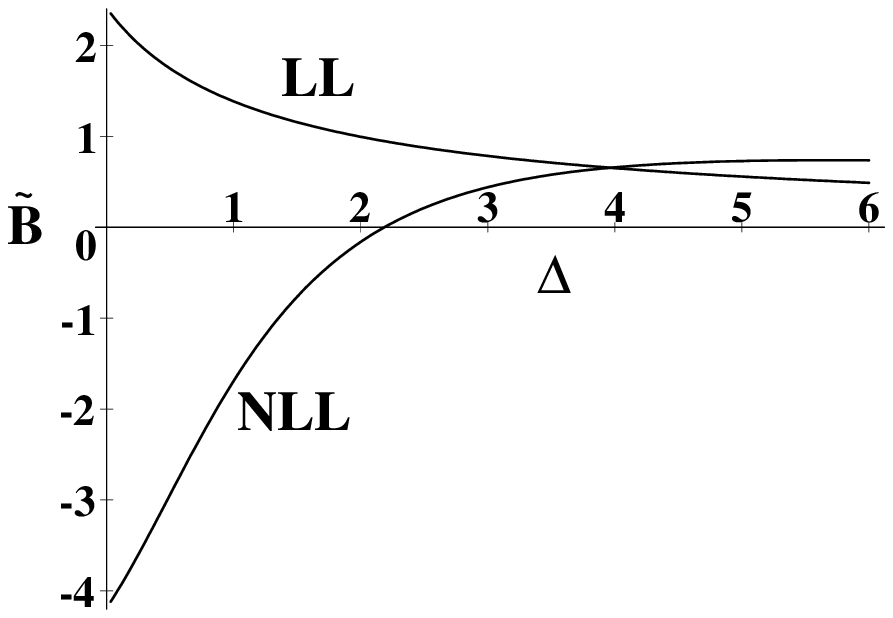}}   
\vskip -.2 cm
\caption[]{
\label{schmidtFigure}
\small Dependence of $\tilde A=\bar\alpha_{s}\chi({1\over2})$ and 
$\tilde B=-{1\over2}\bar\alpha_{s}\chi''({1\over2})$ on $\Delta$ 
for $\alpha_{s}=0.15$, from Ref.~\cite{schmidt}.}
\end{figure}


Figure 3 shows the dependence on $\Delta$ of the leading eigenvalue 
and its second derivative at LL and NLL for $\alpha_{s}=0.15$.
We note that the corrections to $\bar\alpha_{s}\chi({1\over2})$ are not 
large for $\Delta\gtrsim2$ and have weak dependence on $\Delta$ for 
large $\Delta$.  Also, the point $\gamma={1\over2}$ is a maximum for this 
coupling as long as
$\Delta\gtrsim2.2$.  Thus, the BFKL resummation is 
stable for large enough $\Delta$.  We also note that this procedure 
gives predictions of 
$\bar\alpha_{s}\chi({1\over2})$ and $\bar\alpha_{s}\chi''({1\over2})$ for large 
$\Delta$ which are similar to the previous proposal.  However,
the implications of a large value of $\Delta$ for the 
phenomenological use of BFKL is open to interpretation.

Recently, Forshaw, Ross, and Sabio Vera \cite{frv} have studied the inclusion 
of both the double transverse logarithm resummation and the rapidity veto 
simultaneously.  Whereas ref.~\cite{schmidt} emphasized the weak 
dependence on $\Delta$ at large $\Delta$, they were more concerned by 
the strong dependence at small $\Delta$.  They showed 
that after including the resummation of the double transverse 
logarithms, the dependence on the rapidity veto parameter $\Delta$ 
was significantly reduced.  Given the discussion above, this is 
reasonable since both the double transverse logarithm resummation and 
the rapidity veto incorporate the same physical effect: a suppression 
of gluon emissions close by in rapidity.

A third proposal to deal with the large NLL corrections was presented
by Brodsky {\it et al.} \cite{blm}.  They re-evaluated
the NLL corrections in a suitable physical renormalization scheme, and 
then used the BLM procedure \cite{blmone} to find the optimal scale 
setting for the QCD coupling.  The physical connection of this 
proposal to the other two is less obvious; however, as in the previous 
proposals it works by reducing the LL prediction (in this case by the 
choice of the large scale dictated by BLM) combined with a subsequent 
reduction in the NLL corrections.  In addition it yields a very weak 
dependence on the gluon virtuality $p_{1\perp}^{2}$ and leads to an 
approximate conformal invariance.

\section{Conclusions.}

In this talk I have given a brief overview of the BFKL resummation 
program, and have discussed some of the issues that have arisen
from the incorporation of the NLL corrections.  The surprisingly large 
size of the corrections at NLL, as well as the subtle issues related to the 
running of the coupling, have spurred investigations which will lead 
to a better understanding of the physics of QCD at high energies.
Clearly, this is a 
challenging and lively field of theoretical research which will 
significantly impact our understanding of QCD.

\end{document}